\documentclass[%
 superscriptaddress,
 showpacs,
 showkeys,
 nofootinbib,
 amsmath,amssymb,
 aps,
 prd,
 floatfix,
 lengthcheck,%
]{revtex4}
\usepackage{bbding}

\usepackage{graphicx}
\usepackage{dcolumn}
\usepackage{bm}
\usepackage{hyperref}

\usepackage{color}
\usepackage{verbatim}

\newcommand{\ba}{\begin{eqnarray}}
\newcommand{\ea}{\end{eqnarray}}
\newcommand{\be}{\begin{equation}}
\newcommand{\ee}{\end{equation}}
\newcommand {\bp} {{\mathbf p}}
\newcommand {\bx} {{\mathbf x}}
\newcommand {\bq} {{\mathbf q}}
\newcommand {\calL} {{\mathcal L}}
\newcommand {\calO} {{\mathcal O}}

\begin{document}


\title{Radiative transitions in charmonium from $N_f=2$ twisted mass lattice QCD}

\author{Ying Chen}
\affiliation{Institute of High Energy Physics, Chinese Academy of
Sciences, Beijing 100049, P.R. China
}%
\author{De-Chuan Du}
\affiliation{%
School of Physics, Peking University, Beijing 100871, P.~R.~China
}%

\author{Bao-Zhong Guo}
\affiliation{%
School of Physics, Peking University, Beijing 100871, P.~R.~China
}%

\author{Ning Li}
\affiliation{%
School of Physics, Peking University, Beijing 100871, P.~R.~China
}%

\author{Chuan Liu}%
\email[Corresponding author. Email: ]{liuchuan@pku.edu.cn}
\affiliation{%
School of Physics and Center for High Energy Physics, Peking
University, Beijing 100871, P.~R.~China
}%

\author{Hang Liu}
\affiliation{%
School of Physics, Peking University, Beijing 100871, P.~R.~China
}%

\author{Yu-Bin Liu}%
\affiliation{%
Department of Physics, Nankai University, Tianjin, 300071,
P.~R.~China
}%

\author{Jian-Ping Ma}
\affiliation{%
Institute of Theoretical Physics, Chinese Academy of Sciences,
Beijing, 100080, P.~R.~China
}%

\author{Xiang-Fei Meng}
\affiliation{%
Department of Physics, Nankai University, Tianjin, 300071,
P.~R.~China
}%

\author{Zhi-Yuan Niu}
\affiliation{%
School of Physics, Peking University, Beijing 100871, P.~R.~China
}%

\author{Jian-Bo Zhang}
\affiliation{%
Department of Physics, Zhejiang University, Hangzhou, 310027,
P.~R.~China
}%

\collaboration{CLQCD Collaboration}
\noaffiliation
\date{\today}

\begin{abstract}
 We present a study for charmonium radiative transitions:
  $J/\psi\rightarrow\eta_c\gamma$, $\chi_{c0}\rightarrow J/\Psi\gamma$
  and $h_c\rightarrow\eta_c\gamma$ using $N_f=2$
  twisted mass lattice QCD gauge configurations.
  The single-quark vector form factors for $\eta_c$ and $\chi_{c0}$ are
  also determined. The simulation is performed at a lattice spacing of
 $a= 0.06666$~fm and the lattice size is $32^3\times 64$.
 After extrapolation of lattice data
 at nonzero $Q^2$ to $0$~, we compare our results with previous
 quenched lattice results and the available experimental values.
\end{abstract}

 \pacs{12.38.Gc,11.15.Ha}
 \keywords{Charmonium transition, lattice QCD, charmonium form factors.}
\maketitle


\section{Introduction}

 Charmonium physics plays an important and unique role in our
 knowledge of Quantum Chromodynamics (QCD), which is believed to
 be the fundamental theory for strong interactions.
 In some sense, it is comparable to the hydrogen atom
 for atomic physics, the basic theory of which being
 Quantum Electrodynamics (QED). However, charmonium physics is much more involved
 in the sense that, due to its intermediate energy scale
 and the special features of QCD,
 both perturbative and non-perturbative physics
 are present. It is therefore an
 ideal testing ground for our understanding of QCD from both
 perturbative and non-perturbative sides.

 Radiative transitions among various charmonium states
 are particularly important in the study of charmonium physics.
 Most charmonium ground states lie below the open-charm ($D\bar{D}$) threshold
 which makes these states particularly interesting.
 Due to the suppression of the OZI-rule,  these charmonium states usually
 have rather narrow widths. This makes their radiative transitions
 and radiative decays having significant branching ratios
 and are experimentally accessible.
 It is also believed to be the ideal hunting ground for exotic
 hadronic states like the glueballs whose existence is anticipated
 in QCD while its experimental signature remains obscure.
 Recently, the experimental interests have been revived with the
 upgrade for the BESIII experiment at BEPCII storage
 ring~\cite{besiii:instrument-2010,besiii:yellowbook}
 which collects charmonium samples
 that are orders of magnitude larger than ever.

 On the theoretical side, charmonium transitions have been studied using various methods.
 The physical process involves both electromagnetic and
 strong interactions, the former being perturbative in nature while the
 latter being non-perturbative.  Therefore, non-perturbative lattice calculations
 are preferred. Radiative transitions of
 charmonia have been studied comprehensively in quenched lattice QCD for the normal
 ground state charmonia~\cite{dudek:06} and even for some excited and
 exotic ones~\cite{dudek:09}. However, an unquenched lattice study is
 still lacking. In this paper, we would like to pursue
 the feasibility of such a calculation using $N_f=2$ dynamical
 twisted-mass~\cite{Frezzotti01:twistedmass,Frezzotti01:four_quark,Shindler,Boucaud07,Boucaud08,Blossier08,Blossier09,Baron10:charm,Baron10:lightmeson,Alexandrou08,Alexandrou09,Jansen:scalingtest}
 fermion configurations generated by the European Twisted Mass
 Collaboration (ETMC).

 This paper is organized as follows: In Sec.~\ref{sec:method}, we
 briefly describe the lattice setup for the  calculation of the
 hadron matrix element from three-point correlation functions in the
 theory. In Sec.~\ref{sec:simulation_details}, simulation
 details are provided and the results are presented. This
 includes the charmonium spectrum and the dispersion relations,
 the single-quark form factors for $\eta_c$, $\chi_{c0}$ and the radiative transition
 matrix elements responsible for $J/\Psi\to\eta_c\gamma$, $\chi_{c0}\to J/\Psi\gamma$
 and $h_c\to\eta_c\gamma$.
 From these hadronic matrix elements that we obtained in our
 lattice calculation, we compute the transition decay width
 for these channels which are then
 compared with experimental values and the quenched results.
 In Sec.~\ref{sec:conclude} we will summarize our results and conclude.

 \section{Three-point and two-point correlation functions}
 \label{sec:method}

 The lattice setup in this calculation is analogous to the
 the vector form factor calculation of pions which has been studied
 extensively~\cite{Brommel,Hedditch,KFLiu,Boyle07,Boyle08,Capitani,Frezzotti:formfactor}.
 Here we will briefly review the general ideas involved.

 The transitions among charmonium states are triggered by the
 electromagnetic interaction: $\calL^{(e.m.)}_{int}=\int d^4x A^\mu(x)j^{(e.m.)}_\mu(x)$
 between the quark degrees of freedom and the photon field. Here
 $A^\mu(x)$ is the photon field and $j^{(e.m.)}_\mu(x)$ is the electromagnetic vector
 current of the quarks. Since
 the electromagnetic interaction is weak, one can treat it
 perturbatively. This leads to the computation of the
 hadronic matrix element of the current operator
 between the initial ($| i\rangle$) and the final  ($\langle f|$) charmonium states:
 $\langle f|j^{(e.m.)}_\mu(x)|i \rangle$.
 We emphasize that although the electromagnetic interaction
 is perturbative, the matrix element of the current between
 two hadronic states is in general non-perturbative.
 This is the quantity that we would like to compute using
 genuine non-perturbative methods like lattice QCD.

 Within the framework of lattice QCD, charmonium states
 are realized by applying appropriate interpolating
 operators ($\calO_1$ and $\calO_2$ in the formula below)
 to the QCD vacuum $|\Omega\rangle$. Thus, the computation of the hadronic
 matrix element $\langle f|j^{(e.m.)}_\mu(x)|i \rangle$
 naturally leads to the following three-point function:
 \begin{widetext}
 \begin{equation}
 G_\mu(t_2,t;\bp_2,\bp_1)=\sum_{\bx_2,\bx}e^{-i\bp_2\cdot\bx_2}e^{+i{\mathbf q}\cdot\bx}
 \langle\Omega|T\,\mathcal{O}_2(t_2,\bx_2)j^{(e.m.)}_\mu(t,\bx)\mathcal{O}_1^\dagger(0,{\mathbf 0})|\Omega \rangle
 \;.
 \end{equation}
 \end{widetext}
 In this formula, interpolating operators which will
 create/annihilate the appropriate charmonium states are
 inserted at time slices $t=0$ (the source operator)
 and $t=t_2$ (the sink operator), respectively.
 A local operator is used at the source while the
 sink operator with a definite three-momentum $\bp_2$ is utilized.
 The current insertion at time slice $t$ also carries a
 definite three-momentum $\bq$. Momentum conservation then
 implies that the initial state also has a definite
 momentum $\bp_1$ with $\bq=\bp_2-\bp_1$.
 Physically speaking, the three-point function defined above
 represents a process in which an initial charmonium state
 with three-momentum $\bp_1$ created by $\calO^\dagger_1$ making
 a electromagnetic transition to the final charmonium
 state with three-momentum $\bp_2$ annihilated by $\calO_2$
 while the three-momentum difference $\bq$ is
 carried away by the photon.

 Inserting a complete set of states between the electromagnetic
 current operator and the charmonium operators, one finds that,
 when $t_2\gg t \gg 1$, the states with the lowest energy
 dominate the three-point function:
 \begin{widetext}
 \begin{equation}
 \label{eq:threepoint}
 G_\mu(t_2,t;\bp_2,\bp_1)
 \stackrel{t_2\gg t \gg 1}{\longrightarrow}
 \frac{e^{-E_2t_2}e^{-(E_1-E_2)t}}{4E_1(\bp_1)E_2(\bp_2)}
 \langle\Omega|\mathcal{O}_2|f(\bp_2)\rangle
 \langle i(\bp_1)|\mathcal{O}_1^\dagger|\Omega\rangle
 \langle f(\bp_2)|j^{(e.m.)}_\mu(0)|i(\bp_1)\rangle\;.
 \end{equation}
 \end{widetext}
 Therefore, the desired hadronic matrix element
 $\langle f(\bp_2)|j^{(e.m.)}_\mu(0)|i(\bp_1)\rangle$
 can be obtained once the energies $E_1$, $E_2$
 and the corresponding overlap matrix elements
 $\langle\Omega|\mathcal{O}_2|f(\bp_2)\rangle$,
 $\langle i(\bp_1)|\mathcal{O}_1^\dagger|\Omega\rangle$ are known,
 all of which can be obtained from corresponding
 two-point functions for the initial and final
 charmonium states.

 For this purpose, two-point correlation functions
 for the interpolating operators $\calO_i$ for $i=1,2$
 are also computed in the simulation:
 \begin{widetext}
 \be \label{eq:twopoint}
 C_i(t,\bp)\equiv\sum_{\bx}e^{-i\bp\cdot\bx}\langle \Omega| \mathcal
 {O}_i(t,\bx)\mathcal{O}^\dagger_i(0,{\mathbf 0})|\Omega\rangle
 \stackrel{t\gg 1}{\longrightarrow}\frac{|Z_i(\bp)|^2}{E_i(\bp)}e^{-E_i(\bp)\cdot\frac
 {T}{2}}\cosh\left[E_i(\bp)\cdot\left(\frac T2-t\right)\right]\;,
 \ee
 \end{widetext}
 where $Z_i(\bp)=\langle\Omega|\mathcal{O}_i|N(\bp)\rangle$
 is the corresponding overlap matrix element.

 With the relevant two-point and three-point functions,
 the hadronic matrix element $\langle f(\bp_2)|j^{(e.m.)}_\mu(0)|i(\bp_1)\rangle$
 could be extracted using two methods: The first is to fit the
 two-point function Eq.~(\ref{eq:twopoint}) and three-point
 function Eq.~(\ref{eq:threepoint}) simultaneously. The
 second is to form an appropriate ratio from the
 two-point and three-point functions and extract the
 matrix element $\langle f(\bp_2)|j^{(e.m.)}_\mu(0)|i(\bp_1)\rangle$
 directly from the ratio.
 In this study, the second method is utilized and
 the relevant ratio is defined as
 \begin{widetext}
 \begin{eqnarray}
 \label{eq:ratio}
 R_\mu(t)&=&\frac{G_{\mu}(t_2,t;\bp_2,\bp_1)}{C_{2}(t_2,\bp_2)}\sqrt{
   \frac{C_{1}(t_2-t,\bp_1)C_{2}(t,\bp_2)C_{2}(t_2,\bp_2)}
    {C_{2}(t_2-t,\bp_2)C_{1}(t,\bp_1)C_{1}(t_2,\bp_1)  }  }    \nonumber         \\
   &\simeq&\frac{\langle f(\bp_2)|j^{(e.m.)}_\mu(0)|i(\bp_1)\rangle}{4\sqrt{E_2(\bp_2)E_1(\bp_1)}}
 \end{eqnarray}
 \end{widetext}
 where the second line becomes valid when $t_2\gg t \gg 1$,
 assuming only the corresponding ground states dominate.
 In this case, $R_\mu(t)$ becomes independent of $t$ and
 fitting the ratio to a plateau behavior yields the
 desired hadronic matrix element $\langle f(\bp_2)|j^{(e.m.)}_\mu(0)|i(\bp_1)\rangle$.

 Due to different implementations for fermions on the lattice,
 the electromagnetic current operator $j^{(e.m.)}_\mu(x)$ might
 take different forms as compared with its continuum counterpart.
 For Wilson-like fermions, like the twisted mass fermions that we
 use in this study, one could use either the local current or
 the conserved current. The local current is simpler in form but
 it is not conserved on the lattice. It thus
 requires an additional multiplicative renormalization given by
 the factor $Z_V$, which of course can be determined non-perturbatively~\cite{dudek:06}.
 The conserved current is slightly more complicated but due to
 its conservation, it does not need further renormalization, i.e.
 its multiplicative renormalization constant $Z_V\equiv 1$.
 In this work, we use the conserved current and the fact that
 $Z_V=1$ is also verified numerically in our simulation.

 In computing the three-point function defined in
 Eq.~(\ref{eq:threepoint}), various quark contributions arise.
 Since the electromagnetic current consists of contributions
 for all flavors of quarks, light flavors (i.e. $u$, $d$ and $s$ quarks)
 also contribute. Since our charmonium interpolating operators
 are formed only from charm quarks, the contribution from
 the light flavors can only occur through the so-called disconnected
 diagrams. The computation of these diagrams requires the light
 flavor quark propagators at basically all points on the
 lattice (the so-called all-to-all propagators). This is
 computationally extremely costly. Since the total electric
 charge of light quarks adds up to zero, one could argue that
 this contribution vanishes exactly in the flavor $SU(3)$ limit.
 In this study, these contributions are neglected as is
 the case for previous quenched studies~\cite{dudek:06}.
 Thus, we only need the charm quark contribution for
 the electromagnetic current which is proportional to
 the conserved current $j_\mu(x)$ on the lattice via:
 $j^{(e.m.)}_\mu(x)=Q_cj_\mu(x)$
 with $Q_c$ being the electric charge of the charm quark.
 The conserved current $j_\mu(x)$ for the twisted mass
 quark is given by
 \begin{widetext}
 \be \label{eq:current}
 j_\mu(x)=\bar{c}(x)\frac{\gamma_\mu-1}{2}U_\mu(x)c(x+\mu)+\bar{c}(x+\mu)\frac{\gamma_\mu+1}{2}U_\mu^\dagger(x)c(x)
 \ee
 \end{widetext}
 When this current is inserted into the three-point function,
 disconnected diagrams due to the charm quark in principle can
 still can arise. These are neglected in this study since
 charm quark is much heavier than the light quarks and they
 are also OZI-suppressed. Therefore, within the approximations described above,
 we only have to compute the connected diagrams from
 the charm current which can be treated using
 the sequential source method~\cite{sequential:05,dudek:06}.

\section{Simulation details}
\label{sec:simulation_details}

\subsection{The simulation setup for $N_f=2$ twisted mass fermions}

 Twisted mass fermions at the maximal twist are utilized in our study
 with two degenerate light flavors in the sea.
 The framework of maximally twisted mass fermions has
 been utilized in various studies of
 lattice QCD and are shown to be highly promising.
 It offers several advantages when tuned to maximal
 twist: (i) automatic $\mathcal O(a)$ improvement~\cite{Frezzotti01:O(a)improvement} is obtained
 when the bare untwisted quark mass is tuned to its critical value.
 Thus, only one parameter needs to be tuned;
 (ii) The determinant of the twisted mass Dirac operator is strictly positive,
 protecting it against possible zero modes in the so-called exceptional configurations;
 (iii) It simplifies the operator mixing problem for renormalization.

 In this study, gauge field configurations using
 $N_f=2$ ($u$ and $d$ quark) twisted mass fermion are utilized.
 Other quark flavors, namely the strange and charm quarks, are
 introduced as valence quarks.
 As discussed in refs.~\cite{Blossier08,Blossier09,Frezzotti01:four_quark,rehim06},
 we implement non-degenerate valence quarks in the twisted mass formulation
 by formally introducing a twisted doublet for each non-degenerate quark flavor.
 So, in the valence sector we introduce
 three twisted doublets, $(u,d)$, $(s,s')$ and $(c,c')$ with masses
 $\mu_l$, $\mu_s$ and $\mu_c$, respectively. Within each doublet,
 the two valence quarks are regularized in the physical basis
 with Wilson parameters of opposite signs ($r=-r'=1$).
 The fermion action for the valence sector reads:
 \begin{eqnarray}
 S&=&(\bar{\chi}_u,\bar{\chi}_d)\left(D_W+m_{crit}+i\mu_l\gamma_5\tau_3\right)\left({\chi_u \atop \chi_d}\right)\nonumber\\
 & +&(\bar{\chi}_s,\bar{\chi}_{s'})\left(D_W+m_{crit}+i\mu_s\gamma_5\tau_3\right)\left({\chi_s \atop \chi_{s'}}\right)\nonumber\\
 & +&(\bar{\chi}_c,\bar{\chi}_{c'})\left(D_W+m_{crit}+i\mu_c\gamma_5\tau_3\right)\left({\chi_c \atop \chi_{c'}}\right)
 \end{eqnarray}
 where $D_W$ is the usual Wilson-Dirac operator, $m_{crit}$ is the critical
 quark mass which is pre-determined from the simulation by
 ETMC. Numerically, it is more convenient to implement the operators
 in the twisted basis. For this purpose, one can perform a chiral twist and define
 the twisted doublets as follows:
 \begin{eqnarray}
 \left({u \atop d}\right )&=&\exp\left(i\omega\gamma_5\tau_3/2\right)\left({\chi_u \atop \chi_d}\right)\nonumber\\
 \left({s \atop s'}\right )&=&\exp\left(i\omega\gamma_5\tau_3/2\right)\left({\chi_s \atop \chi_{s'}}\right)\nonumber\\
 \left({c \atop c'}\right )&=&\exp\left(i\omega\gamma_5\tau_3/2\right)\left({\chi_c \atop \chi_{c'}}\right)
 \end{eqnarray}
 where $\omega=\pi/2$ implements the full twist.

 In this work, all computations are done using $N_f=2$ twisted
 mass fermion configurations at the lattice spacing of
 $a=0.0666$fm ($\beta=4.05$). The size of the lattice is $32^3\times 64$
 so that the spatial extent of the lattice is about $2.13$fm,
 which is a safe value for charmonium physics.
 In the temporal direction, anti-periodic boundary condition
 is applied for the quark field
 while periodic boundary condition
 is utilized in all spatial directions.
 The simulation parameters for our study are summarized
 in Table ~\ref{tab:lattice} .
 \begin{table}[ht]
 \centering \caption{Simulation parameters in this study. }
 \label{tab:lattice}
 \begin{tabular*}{8cm}{@{\extracolsep{\fill}}|c|c|c|c|c|c|c|}
 \hline
 $L^3*T $   &$\beta$   &$ \kappa_{c} $    &$a$\,[fm]   &$ a\mu$    &$m_{\pi}$\,[MeV]   &$N_{\rm conf}$ \\
 \hline
 $32^3*64$   &4.05    &0.15701        &0.0666    &0.0080      &488           & 201  \\
 \hline
 \end{tabular*}
 \end{table}

 As for the charmonium states, we build the interpolating field
 operators within the same Wilson parameters. In the physical basis, they read $\bar{c}\Gamma c$
 and the corresponding form in twisted basis $\bar{\chi}_c\Gamma'\chi_c$ can also be obtained
 easily. These are tabulated in Table ~\ref{tab:operator} together
 with the  possible $J^{PC}$ quantum numbers in the continuum
 and the names of the corresponding particle.
 \footnote{In the first row of the table, we list the names of
 the charmonium states where $\chi_{c0}$ and $\chi_{c1}$ are
 not to be confused with the charm quark field in the
 twisted basis.}
 \begin{table}[h]
 \centering \caption{Local interpolating fields for charmonium states
 studied in this work in both physical and twisted basis, $\bar{c}\Gamma
 c=\bar{\chi}_c\Gamma'\chi_c$. Also listed are
 the names of the corresponding particle and their
 $J^{PC}$ quantum numbers in the continuum.} \label{tab:operator}
 \begin{tabular*}{8cm}{@{\extracolsep{\fill}}cccccc}
 \hline\hline
                  &$J/\psi$        &$\eta_c$     &$\chi_{c0}$     &$\chi_{c1}$        &$h_c$\\[1mm]\hline
 $\Gamma$          &$\gamma_i$     &$\gamma_5$      &$1$           &$\gamma_i\gamma_5$  &$\sigma_{ij}$\\[2mm]
 $\Gamma'$         &$\gamma_i$     &$1$             &$\gamma_5$    &$\gamma_i\gamma_5$  &$\sigma_{0i}$\\[2mm]
 $J^{PC}$          &$1^{--}$       &$0^{-+}$        &$0^{++}$      &$1^{++}$            &$1^{+-}$\\[2mm]\hline
 \end{tabular*}
 \end{table}

 Two-point functions are computed as usual for all charmonium states
 involved (those listed in Table~\ref{tab:operator}) in our calculation.
 Fitting these two-point functions yields
 the energy for the corresponding charmonium states, both with and
 without three-momentum. As for the three-point functions,
 since only connected diagrams involving charm propagators are
 needed, sequential source method is utilized~\cite{sequential:05}.
 The results for the two-point and three-point functions are
 then employed to construct the relevant ratio
 defined in Eq.~(\ref{eq:ratio}).
 For definiteness, we set $t_2=32$ in our simulations which makes the three point
 function anti-symmetric (for $j_{\mu=0}$) or symmetric
 (for $j_{i}$ with $i=1,2,3$)~\cite{Brommel}
 about the time slice $t_2=32$. In practice, we average the data from the
 two halves to improve statistics.
 All errors in this study are estimated using the
 conventional jack-knife method.

 \subsection{Charmonium spectrum and  dispersion relations}
 \label{subsec:masses}

 Before computing the transition matrix element, the mass and
 the energy dispersion relations for the relevant charmonium states
 have to be verified. This is particularly important for our study
 due to the following reasons. Although charmonium spectrum has been
 studied extensively in quenched lattice QCD and the overall picture
 agrees reasonably well with the experiment, some quantities like
 the mass splitting between $\eta_c$ and $J/\Psi$ disagrees with
 the experimental value. It is widely believed that this discrepancy mainly
 originates from the quenched approximation. It is therefore useful
 to check, using unquenched twisted mass configurations, whether this
 discrepancy can be resolved. Furthermore, although twisted mass
 configurations have been used successfully to study the light
 flavors, using them on heavy charm quark needs some care.
 Being relatively heavy, the charm quark mass parameter $\mu_ca\sim 0.2$
 in our study is not tiny. Of course, the good news from the maximally twisted mass
 lattice QCD is that it is $\calO(a)$ improved.
 Therefore, one would still hope to bring the lattice discretization errors
 under control. Another measure of the possible
 lattice artifacts is the charmonium mass, say the mass
 of the $\eta_c$ meson $m_{\eta_c}$ in lattice unit.
 In our study, it turns out that $m_{\eta_c}a\sim 1$.
 For charmonium states with non-zero three-momentum,
 this number becomes even larger.
 Therefore, one should carefully verify that these
 possible lattice artifacts for the charmonium states are not
 out of control. Only after these reassurances can one
 possibly proceed to calculate transitions among charmonium
 states reliably. As we will illustrate below, in our simulation, most
 of these lattice artifacts are remedied by using the
 lattice dispersion relations for the charmonium states.

 Following Eq.~(\ref{eq:twopoint}), the energy
 $E(\bp)$ for a particular charmonium state
 with three-momentum $\bp$ can be obtained from
 the corresponding two-point function via
 \be
 \cosh\left(E(\bp)\right)=\frac{C(\bp;t-1)+C(\bp;t+1)}{2C(\bp;t)}
 \ee
 The two point function is symmetric about $t=T/2$. In real simulation
 we average the data from two halves about $t=T/2$ to improve statistics.
 For each channel, several three-momenta (including the zero three-momentum)
 have been computed. Different momentum modes that are related by
 lattice symmetries are averaged over.

 \begin{figure*}[h]
 \centering
 \rotatebox{-90}{\includegraphics[scale=0.60]{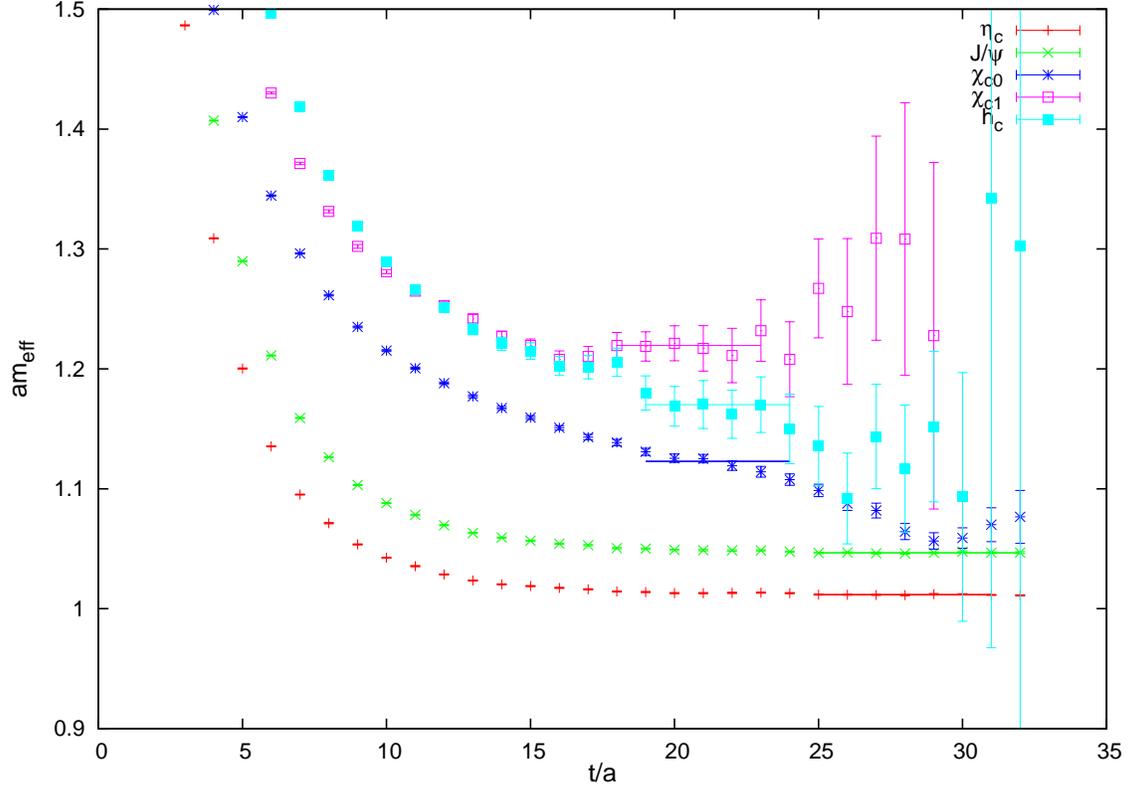}}
 \caption{Charmonium effective mass plateaus. From lower to higher
 values, the plateau corresponds to the charmonium state
 $\eta_c$, $J/\Psi$, $\chi_{c0}$, $h_c$ and $\chi_{c1}$, respectively}
 \label{fig:massplateau}
 \end{figure*}
 The effective mass plateaus at zero three-momentum for the charmonium states
 studied in this work are illustrated in Fig.~\ref{fig:massplateau}.
 From lower to higher
 values, the plateau corresponds to the charmonium state of
 $\eta_c$, $J/\Psi$, $\chi_{c0}$, $h_c$ and $\chi_{c1}$, respectively.
 It is seen that the effective mass values for $\eta_c$ and $J/\Psi$ have
 shown very clear and well-established plateau behavior,
 resulting in rather small statistical errors.
 We use the mass of $J/\Psi$ from our simulation to set the bare charm
 quark mass parameter $\mu_c$. After some tuning, we find
 $a\mu_c\simeq 0.203$ roughly corresponds to the value
 that is consistent with the value quoted in the Particle Data Group (PDG).
 We then fix $\mu_c$ at this particular value for all our subsequent calculations.
 Since twisted mass lattice QCD is $\calO(a)$ improved,
 the anticipated cutoff effects induced by the charm quark mass
 is roughly $\calO(a^2\mu_c^2)$, which is at a few percent level.
 This of course still
 needs further verification from measured physical quantities.
 The effective mass plateaus for other charmonium states: $\chi_{c0}$,
 $\chi_{c1}$ and $h_c$ are relatively noisy with larger statistical errors.
 The fitted effective mass values are collected in Table ~\ref{tab:charmmass}
 which can be compared with the corresponding values from PDG.

 \begin{table}[h]
 \centering \caption{Charmonium effective mass~ [Unit:MeV]}
 \label{tab:charmmass}
 \begin{tabular*}{8cm}{@{\extracolsep{\fill}}|c|c|c|c|c|c|}
 \hline
          &$\eta_c$      &$J/\psi$      &$\chi_{c0}$      &$\chi_{c1}$      &$h_c$\\ \hline
 Mass (this work)   &2997.4       &3101.0         &3326.9           &3613.2           &3466.8\\ \hline
 Error & 0.5         &0.7            &4.5             &18.4             &23.0     \\ \hline
 PDG       &2980.3       &3096.9         &3414.7           &3510.7           &3525.9\\ \hline
 \end{tabular*}
 \end{table}
 It is gratifying to see that our lattice result suggests
 $m_{J/\Psi}-m_{\eta_c}=104$MeV,
 a little smaller than the PDG value of about $117$ MeV.
 This is already a great improvement over the quenched studies
 where the lattice results are typically away by dozens of MeV.
 This remaining discrepancy might come from lattice
 artifacts (since we are simulating at a fixed lattice
 spacing without taking the continuum extrapolation)
 and/or from the fact that we have neglected
 annihilation diagrams for the charm quark
 in the two-point function, as estimated in Ref.~\cite{detar:10}.

 \begin{figure*}[ht]
 \centering
 \includegraphics[scale=0.8]{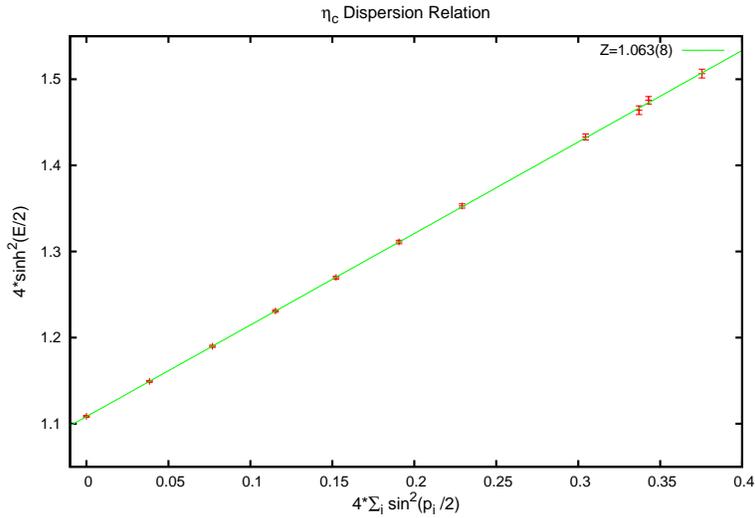}
 \caption{The $\eta_c$ dispersion relation obtained from our
 calculation. Following the lattice dispersion
 relation~(\ref{eq:lattice_disper}),
 the quantity $4\sinh^2\left( \frac {E(\bp)}{2}\right)$ (vertical axis) is plotted
 versus different values of $4\sum_i\sin^2\left(\frac {p_i} 2\right)$ (horizontal axis).
 The data points with errors are simulation results while the straight
 line is a linear fit according to Eq.~(\ref{eq:lattice_disper}) with
 the fitted value of $Z$ indicated in the upper right corner of the plot. } \label{fig:etac_disper}
 \end{figure*}
 To get a feeling about the size of the
 lattice artifacts for the charmonium states with
 non-vanishing three-momenta, we investigate the
 dispersion relations for $\eta_c$, $J/\Psi$ and $\chi_{c0}$ states.
 The energy $E(\bp)$ are obtained from the corresponding effective
 mass plateaus of the two-point functions with prescribed three-momentum.
 As said in the beginning of this subsection, since the charmonium
 states are relatively heavy in lattice unit,
 the continuum dispersion relation $E^2=m^2+c^2\bp^2$ may not
 be a good description, where $c$ is the speed of light which
 should be close to unity if lattice artifacts are small.
 Indeed, our data suggest that the naive continuum dispersion relation is violated
 with the fitted value of $c^2$ substantially away from unity
 by as much as 12\% even for $\eta_c$ and $J/\Psi$ states.
 However, we find that, if we utilize the standard
 lattice dispersion relation
 \be
 \label{eq:lattice_disper}
 4\sinh^2\left( \frac {E(\bp)}{2}\right) =4\sinh^2 \left(\frac m2\right)
 +Z\times 4\sum_i\sin^2\left(\frac {p_i} 2\right)\;,
 \ee
 which recovers the naive dispersion relation in
 the continuum limit, we could describe our data extremely well with
 the fitted values of $Z$ for $\eta_c$ and $J/\Psi$ rather close to unity.

 \begin{figure*}[ht]
 \centering
 \includegraphics[scale=0.8]{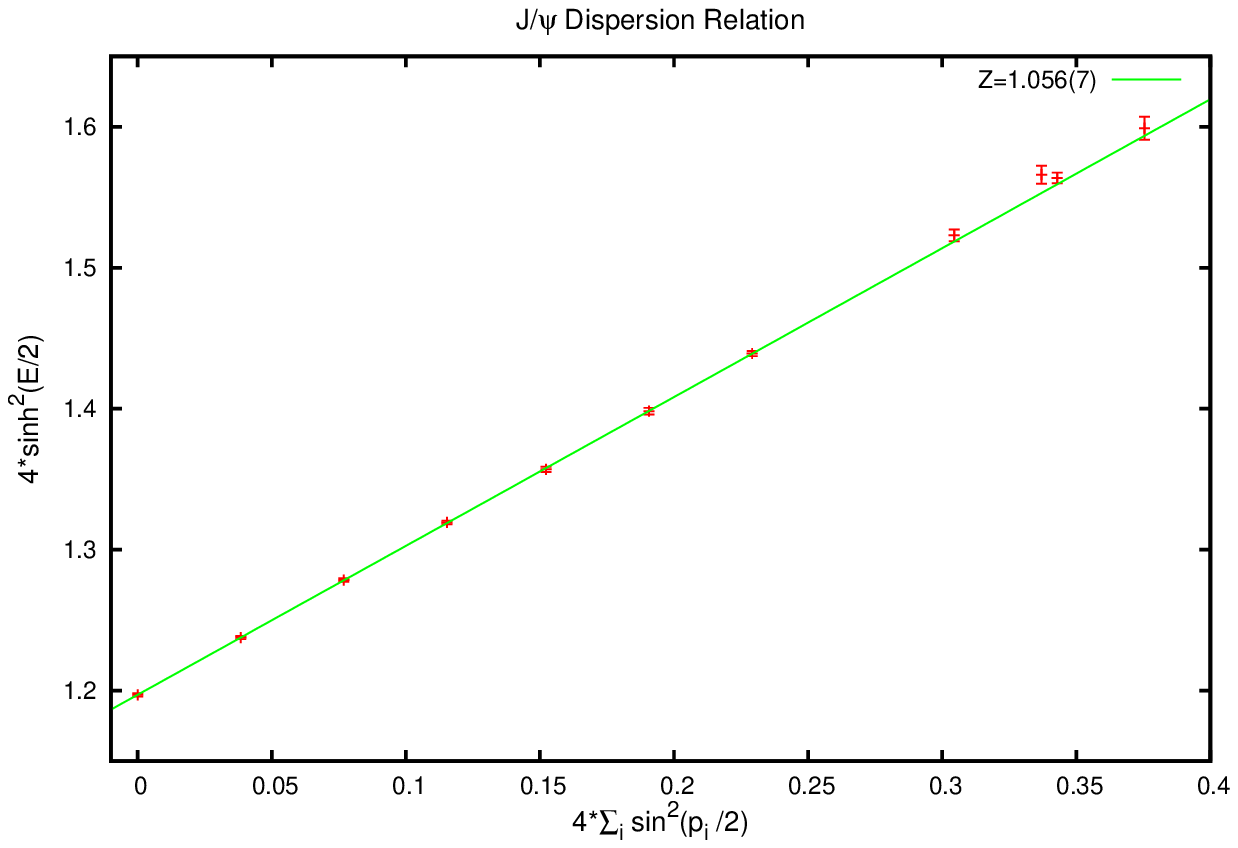}
 \caption{The same as Fig.~\ref{fig:etac_disper} but
 for $J/\Psi$.} \label{fig:jpsi_disper}
 \end{figure*}
\begin{figure*}[h]
 \centering
 \includegraphics[scale=0.8]{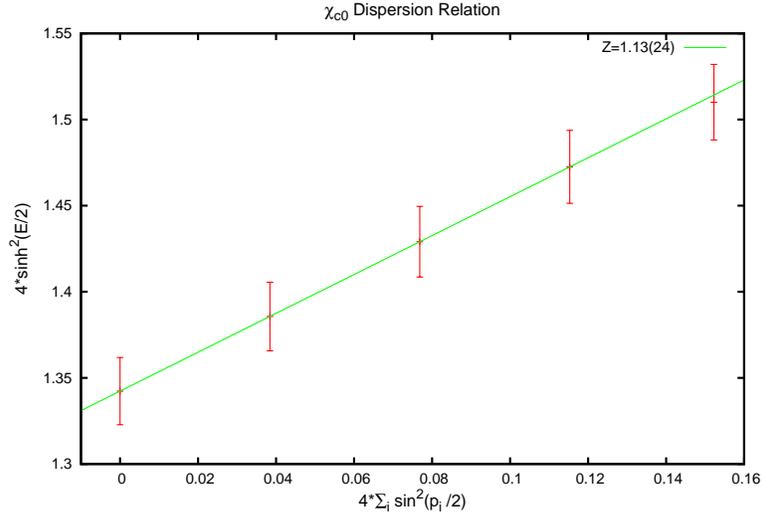}
 \caption{The same as Fig.~\ref{fig:etac_disper} but
 for  $\chi_{c0}$.}
 \label{fig:chic0_disper}
 \end{figure*}

 The dispersion relations for $\eta_c$, $J/\Psi$ and $\chi_{c0}$
 are illustrated in
 Fig.~\ref{fig:etac_disper},
 Fig.~\ref{fig:jpsi_disper} and
 Fig.~\ref{fig:chic0_disper}, respectively.
 We find that $Z_{\eta_c}=1.063(8), Z_{J/\Psi}=1.056(7), Z_{\chi_{c0}}=1.13(24)$, all
 of which are close to the anticipated value $Z\equiv 1$.
 The difference seems to be at the order of $\calO((\mu_ca)^2)\sim
 4\%$ as the naive estimate suggests.
 In evaluating the two-point function, thanks to the averaging of various $\bp$
 related by lattice symmetries, we get very good dispersion relation even at $\bp^2$
 as large as $10\bp_{min}^2$, where $\bp_{min}=(100)$ (in unit of
 $(2\pi)/L$) is the minimal lattice momentum.
 It is also seen that even at the largest three-momentum, the
 lattice dispersion relation still offers a very good description of
 the data. This gives us confidence that, at this particular lattice spacing that
 we are simulating, most of the lattice
 artifacts for the charmonium states are taken care of
 by using the lattice dispersion relation~(\ref{eq:lattice_disper}).

\subsection{Form factors for $\eta_c$ and $\chi_{c0}$}

 In the continuum, the hadronic matrix element
 $\langle \eta_c(\bp_2)|j_\mu(0)|\eta_c(\bp_1)\rangle$ may be
 parameterized by only one form factor $f(Q^2)$ as:~\cite{dudek:06}
 \be
 \label{eq:etac_decomposition}
 \langle \eta_c(\bp_2)|j_\mu(0)|\eta_c(\bp_1)\rangle
 \equiv f(Q^2)(p_1+p_2)_{\mu}\;,
 \ee
 where $Q^2\equiv -(p_2-p_1)^2$ is the square of four momentum transfer.
 This quantity is also called the single-quark elastic form factor
 in Ref.~\cite{Lakhina:2006vg}. It is not a directly measurable
 quantity experimentally. But it is a quantity that can be computed
 in lattice simulations which can then be utilized to compare
 with similar results from models (see Ref.~\cite{Lakhina:2006vg}).
 Note that this form automatically ensures the current conservation
 $\langle \eta_c(\bp_2)|\partial^\mu j_\mu|\eta_c(\bp_1)\rangle=0$
 since $q\cdot(p_1+p_2)=(p_2-p_1)\cdot(p_2+p_1)=0$.
 On a finite lattice, the partial derivatives are replaced
 by corresponding finite differences on the lattice.
 For the temporal components of the four-momenta,
 this amounts to replacing the continuum energy
 by its lattice counter part:
 $(p_i)_0 \rightarrow 2\sinh(E_i/2)$. Note
 that this modification applies to
 the energy factors inside the square root in
 the second line of Eq.~(\ref{eq:ratio}).
 Of course, in principle the spatial components should also
 be modified according to the lattice
 dispersion relation~(\ref{eq:lattice_disper}).
 But since our three-momenta are relatively small
 in lattice unit, this replacement does not
 make a significant change. For the temporal components,
 however, since $aE(\bp)\sim 1$ for all charmonium states
 being studied, this modification is crucial.
 For example, according to Eq.~(\ref{eq:etac_decomposition}),
 the form factor $f(Q^2)$ is a scalar function which
 is the same for all indices $\mu=0,1,2,3$.
 Only after using the modifications suggested by
 the lattice dispersion relation can we obtain
 consistent results for $f(Q^2)$ at
 different values $\mu$.

 \begin{figure*}[ht]
 \includegraphics[scale=0.8]{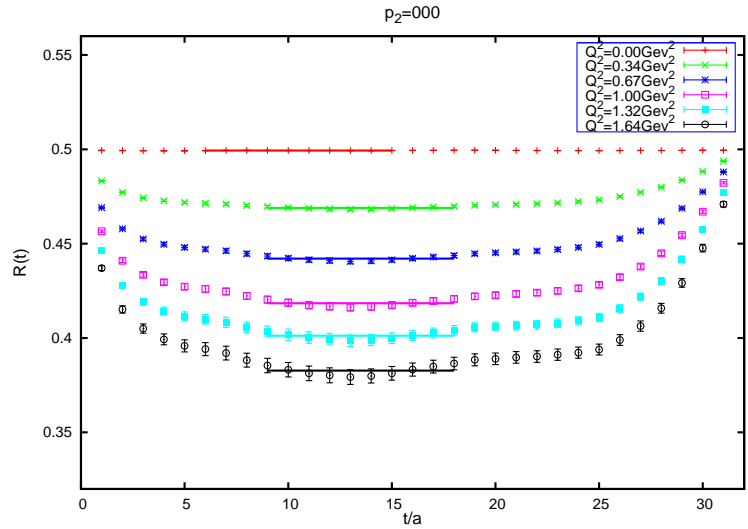}
 \caption{The ratio $R_0(t)$ defined in Eq.~(\ref{eq:ratio}) for
 $\eta_c$ with $\bp_2=(0,0,0)$.}
 \label{fig:etac_ratio_pf000}
 \end{figure*}
\begin{figure*}[ht]
 \includegraphics[scale=0.8]{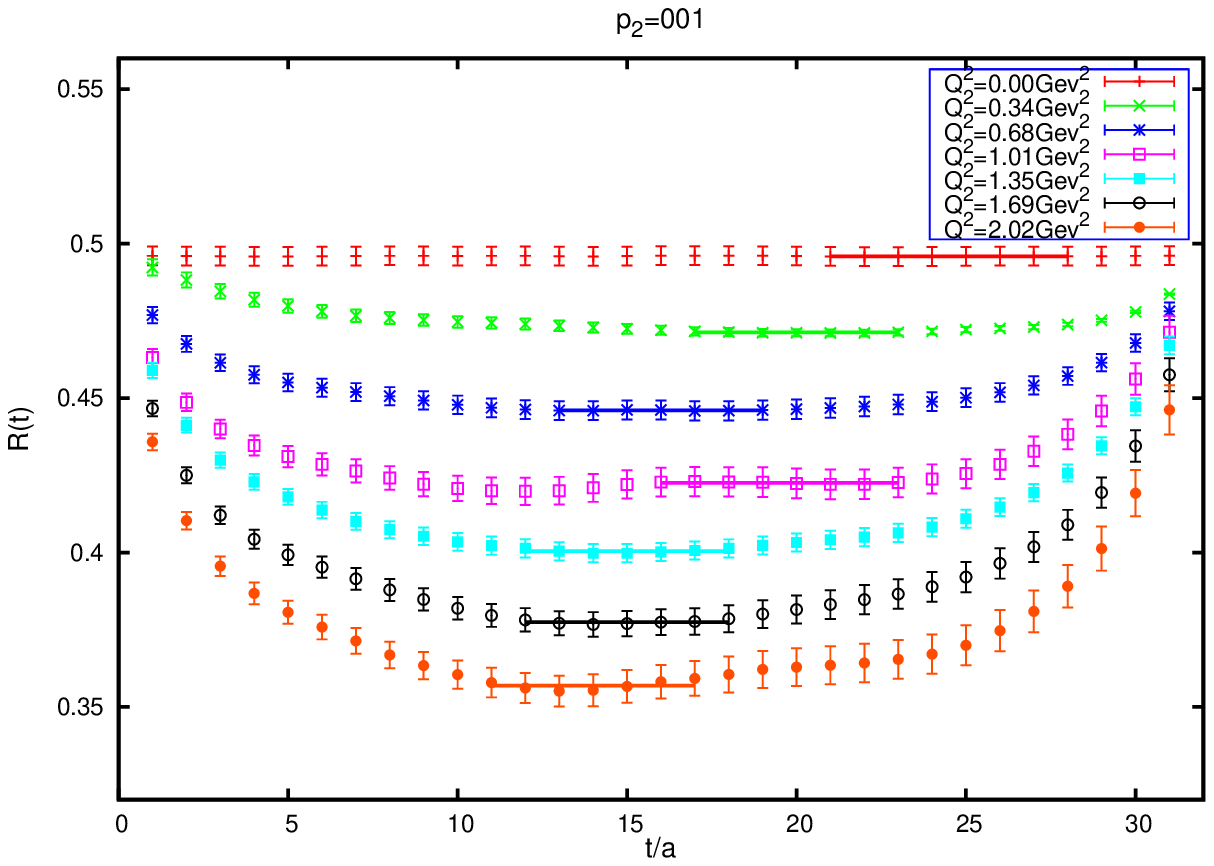}
 \caption{The ratio $R_0(t)$ defined in Eq.~(\ref{eq:ratio}) for
 $\eta_c$  with $\bp_2=(0,0,1)$.}
 \label{fig:etac_ratio_pf001}
 \end{figure*}

 To obtain the desired hadronic matrix element $\langle \eta_c(\bp_2)|j_\mu(0)|\eta_c(\bp_1)\rangle$,
 we form the ratio defined in Eq.~(\ref{eq:ratio}).
 This is done for the zero three-momentum case $\bp_2=(0,0,0)$ and
 for various non-vanishing three-momenta.
 In Fig.~\ref{fig:etac_ratio_pf000} and Fig.~\ref{fig:etac_ratio_pf001},
 we display the typical behaviors for $R_0(t)$ for
 $\bp_2=(0,0,0)$ and $\bp_2=(0,0,1)$, respectively.
 It is seen that clear plateau behaviors have been established
 from which the form factor $f(Q^2)$ can be extracted. We have
 checked that taking the temporal and spatial components of
 the current yields consistent results for $f(Q^2)$ although
 the results obtained from $\mu=0$ (i.e. $R_0(t)$) gives smaller statistical
 errors, which we take as the final result for the
 form factor at that particular $Q^2$.

\begin{figure*}[ht]
 \centering
 \includegraphics[scale=0.8]{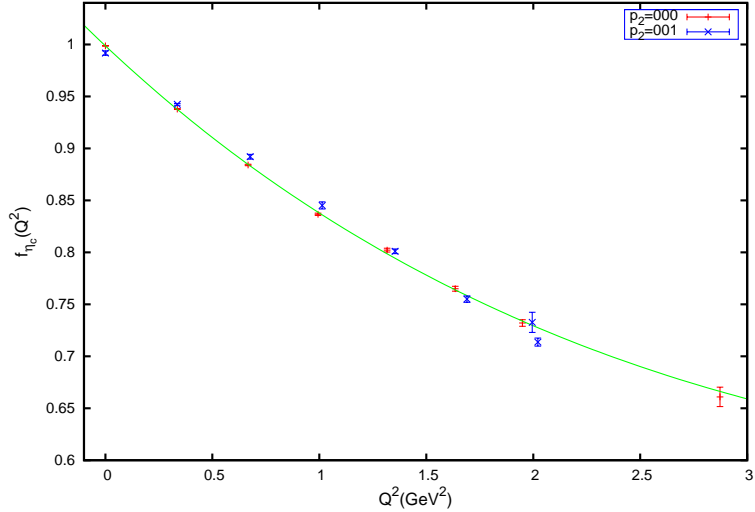}
 \caption{The form factor $f(Q^2)$ for $\eta_{c}$ obtained
 from $\bp_2=(0,0,0)$ (red data points) and
 $\bp_2=(0,0,1)$ (blue data points). The curve
 is a fit for all the data using the functional
 form of Eq.~(\ref{eq:etac_form_fit}). } \label{fig:etac_form_factor}
 \end{figure*}

 The fitted values of $f(Q^2)$ obtained from the ratio
 are shown in Fig.~\ref{fig:etac_form_factor} versus different
 values of $Q^2$
 where two different type of symbols stands for
 $\bp_2=(0,0,0)$ and $\bp_2=(0,0,1)$, respectively.
 It is seen that the data obtained in the  two
 cases tend to lie on a universal curve.
 Following Ref.~\cite{dudek:06}, we fit the data for the
 form factor with the following function:
 \be
 \label{eq:etac_form_fit}
 f(Q^2)=\exp\left[-\frac{Q^2}{16\beta^2}\left(1+\alpha Q^2\right)\right]
 \ee
 The fitted parameters turn out to be:
 \be
 \alpha= -0.096(6)~ \mbox{GeV}^{-2},\;  \beta = 567(2)~ \mbox{MeV}
 \ee
 This value of $\beta$ is larger than the corresponding value
 $480(3)$MeV obtained in the quenched approximation in
 Ref.~\cite{dudek:06}, making the corresponding form factor
 obtained from our unquenched calculation ``harder''
 (i.e. decays slower with increasing $Q^2$).
 The comparison of this form factor obtained from various phenomenological models
 with the corresponding quenched result has been addressed in
 Ref.~\cite{dudek:06,Lakhina:2006vg}.
 It was noted that using the simple harmonic oscillator (SHO)
 wavefunctions yields a harder form factor when compared
 with the quenched lattice result. In the quark model
 of ISGW~\cite{Isgur:1988gb}, however, an extra factor $\kappa\simeq 0.7$ was
 introduced such that the form factor takes the
 form $f(Q^2)\sim \exp\left(-Q^2/(16\beta^2\kappa^2)\right)$ near $Q^2=0$ which
 agrees with the {\em quenched} lattice result well, given a phenomenological
 value of $\beta\sim 710$MeV.
 Since our unquenched lattice result suggests a harder
 behavior for the form factor than the quenched case, we find that, for
 the same value of $\beta$ taken in the model, a factor
 of $\kappa\simeq 0.8$ will make the model predictions in
 good agreement with our unquenched lattice results.
 One can define a squared mean charge radius $\sqrt{\langle r^2 \rangle}$
 with $\langle r^2 \rangle$ given by:
 \be
 \langle r^2 \rangle=-6\left.\frac{d}{dQ^2}f(Q^2)\right|_{Q^2=0}
 =\frac{6}{16\beta^2}
 \;.
 \ee
 Our unquenched lattice result then yields $\sqrt{\langle r^2 \rangle}=0.213(1)~\mbox{fm}$
 which is smaller than the corresponding quenched value of $0.255(2)$fm.

 Our unquenched result yields a harder behavior for the form factors
 which can be understood qualitatively. Physical  scales on the lattice are
 usually set by some long-distance physical quantities, like the
 static quark anti-quark potential in the quenched or the pion decay constant in
 the $N_f=2$ twisted mass lattice QCD. However, it is known that quenched lattice
 QCD did not reproduce the true QCD $\beta$-function due to the lack
 of the quark loops. In particular, when running from the lower
 energy scale up to the scale of charmonium physics, quenched
 lattice QCD gives a weakened strong coupling constant than
 unquenched lattice QCD. This is believed to be the major reason for the
 discrepancy between the mass splitting of $J/\Psi$ and $\eta_c$
 in quenched lattice QCD with the true experimental result.
 Therefore, unquenching the quarks will basically make
 the effective coupling constant stronger at charmonium scale when
 compared with the quenched case. This in turn gives a smaller
 charge radius for the unquenched case, in agreement with what we find
 in our calculation.

\begin{figure*}[ht]
 \centering
 \includegraphics[scale=0.8]{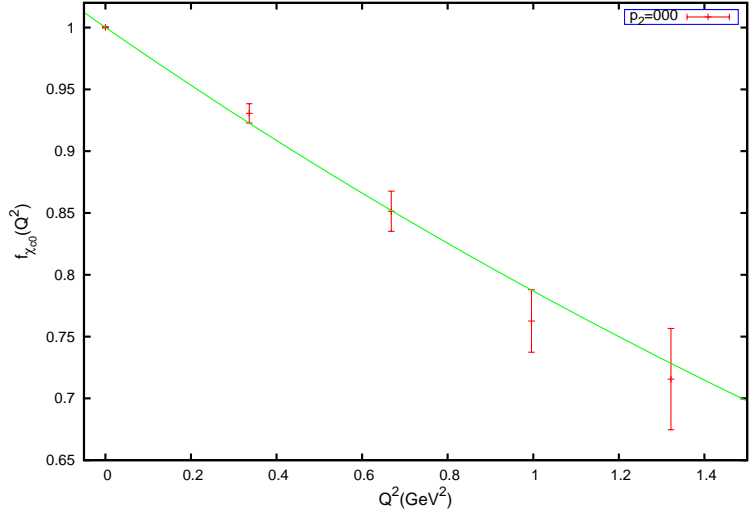}
 \caption{The same as Fig.~\ref{fig:etac_form_factor} except that
 the charmonium state is $\chi_{c0}$.} \label{fig:chic0_form_factor}
 \end{figure*}

 The hadronic matrix element $\langle \chi_{c0}(\bp_2)|j_\mu(0)|\chi_{c0}(\bp_1)\rangle$
 for $\chi_{c0}$ has the same form of decomposition as that for $\eta_c$.
 The corresponding  form factor is
 defined as in Eq.~(\ref{eq:etac_decomposition}).
 In exactly the same manner, we can obtain the form factor
 $f(Q^2)$ for $\chi_{c0}$ except that we have only
 computed the case $\bp_2=(0,0,0)$.
 This is illustrated in Fig.~\ref{fig:chic0_form_factor}.
 The data is fitted with the function:
 \be
 f(Q^2)=f(0)\exp\left[-\frac{Q^2}{16\beta^2}\right]
 \ee
 The fit parameters are:
 \be
 f(0)=1.0002(5),\;\beta=510(16)~\mbox{MeV}
 \ee
 This value of $\beta$ is also larger than
 the quenched value of $393(12)$MeV from Ref.~\cite{dudek:06}
 (i.e. the unquenched form factor is also harder than the quench one).
 Note that our fitted value of $\beta$ for $\chi_{c0}$ is
 smaller than that for $\eta_c$ resulting in a larger
 charge radius for $\chi_{c0}$ when compared with that
 for $\eta_c$. This is consistent with the quark model
 picture since $\chi_{c0}$ in this model is a $L=1$
 state and the charge radius is naturally larger due to
 the presence of the centrifugal potential.

 Finally, we remark that, in cases of both $\eta_c$ and $\chi_{c0}$,
 although some data might have large errors at non-vanishing $Q^2$,
 the form factor as zero momentum transfer is always consistent with
 unity which is in fact a manifestation of the current conservation:
 $Z_V=1$. Therefore, the conserved current requires no extra
 multiplicative renormalization as it should. This is verified
 numerically by our simulation data.

\subsection{$J/\Psi\to \eta_c\gamma$ transition}

\begin{figure*}[ht]
 \centering
 \includegraphics[scale=0.8]{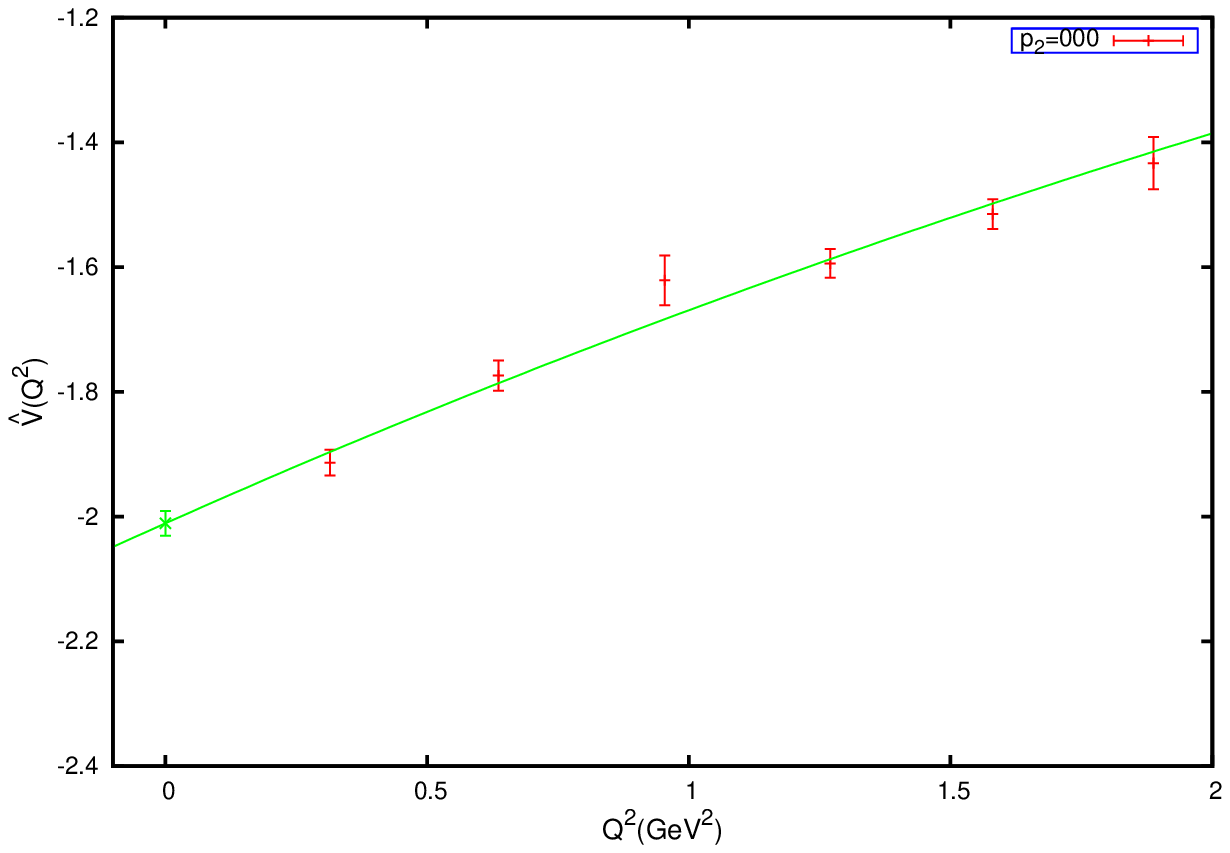}
 \caption{The lattice results for the $J/\Psi \to \eta_c\gamma$ transition
 form factor $\hat{V}(Q^2)$. The curve is a fit using
 the function in Eq.~(\ref{eq:jpsi-to-etac-fit}). The fitted value
 of $\hat{V}(Q^2=0)$ is also shown at $Q^2=0$ together
 with its corresponding error.}
 \label{fig:jpsi_to_etac}
 \end{figure*}

 The matrix element $\langle \eta_c(\bp_2)|j^\mu(0)|\left[J/\Psi\right]_r(\bp_1)\rangle$
 is responsible for the calculation of  $J/\Psi\to \eta_c\gamma$ transition rate.
 Here we use the index $r$ to designate the polarization of
 the initial $J/\Psi$ state whose polarization vector
 is denoted by $\epsilon_\gamma(\bp_1,r)$.
 In the continuum, this matrix element can be decomposed
 as~\cite{dudek:06}
 \begin{widetext}
 \be
 \label{eq:jpsi_decomposition}
 \langle \eta_c(\bp_2)|j^\mu(0)|\left[J/\Psi\right]_r(\bp_1)\rangle\equiv
 \frac{2V(Q^2)}{m_{\eta_c}+m_\Psi}\epsilon^{\mu\alpha\beta\gamma}p_{2\alpha}p_{1\beta}
 \epsilon_\gamma(\bp_1,r)
 \;,
 \ee
 \end{widetext}
 Thus the matrix element is characterized by one form factor
 $V(Q^2)$. By forming the appropriate ratio, relevant lattice results
 $\hat{V}(Q^2)$ are extracted from the
 plateaus of the ratios. The relation of $\hat{V}(Q^2)$ with its continuum counterpart
 $V(Q^2)$ is $V(Q^2)=2\times \frac23e\times\hat{V}(Q^2)$, where
 the factor $2$ comes from the quark and the anti-quark while
 the factor $(2e/3)$ is due to the charge of the charm quark.
 The results for the transition form factor $\hat{V}(Q^2)$ thus obtained
 are illustrated in Fig.~\ref{fig:jpsi_to_etac}. Following Ref.~\cite{dudek:06},
 the data is fitted with the function:
 \be
 \label{eq:jpsi-to-etac-fit}
 \hat{V}(Q^2)=\hat{V}(0)\exp\left[-\frac{Q^2}{16\beta^2}\right]
 \;.
 \ee
 The resulting fitted parameters we find are as follows:
 \be
 \hat{V}(0)= -2.01(2)   ,\,    \beta = 580(19)~\mbox{MeV}
 \ee
 This is to be compared with similar results from
 previous quenched lattice study: $\hat{V}(0)= -1.85(4)$ and $\beta = 540(10)~\mbox{MeV}$
 in Ref.~\cite{dudek:06}.

 With the values of the transition form factor on the lattice,
 the $J/\Psi \to \eta_c\gamma$ decay width can be obtained:
 \be
 \label{eq:jpsi_decay_width}
 \Gamma\left(J/\Psi \to \eta_c\gamma\right)=\alpha\frac{64}{27}
 \frac{|\bq|^3}{(m_{\eta_c}+m_\Psi)^2}|\hat{V}(0)|^2
 \ee
 where $\bq$ is frame dependent. If we choose the frame in which the initial $J/\Psi$ is
 at rest, we have:
 $|\bq|^2=\left(m_\Psi^2-m_{\eta_c}^2\right)^2/(4m_\Psi^2)$.
 Substitute this into Eq.~(\ref{eq:jpsi_decay_width})
 we then get the $J/\Psi \to \eta_c\gamma$ decay width:
 \be
 \Gamma_{m_{\rm phy}}=2.84(6)~\mbox{KeV},\,\Gamma_{m_{\rm lat}}=1.99(6)~\mbox{KeV}
 \ee
 where $\Gamma_{m_{\rm phy}}$ denotes the result with physical mass
 values (e.g. values from PDG)
 are substituted into Eq.~(\ref{eq:jpsi_decay_width}),
 while $\Gamma_{m_{\rm lat}}$ stands for using the mass values
 computed from the lattice directly.
 This difference arises since our lattice results for
 the masses for $J/\Psi$ and $\eta_c$ in subsection~\ref{subsec:masses}
 (see Table~\ref{tab:charmmass}) do not coincide
 with their experimental values exactly.
 Although our lattice value for the mass of the $J/\Psi$ is quite
 close and that for $\eta_c$ is also closer to the experimental
 value than the corresponding quenched value, the decay width
 turns out to be proportional to $(m_{J/\Psi}-m_{\eta_c})^3$ which
 magnifies the difference.
 Note that, for quenched lattice calculations, using different
 charmonium mass values makes a even bigger difference, as noted
 in Ref.~\cite{dudek:06}. The corresponding results are:
 $\Gamma_{m_{\rm phy}}=2.57(11)$ KeV,
 $\Gamma_{m_{\rm lat}}=1.61(7)$ KeV.
 This is due to the fact that quenched lattice calculations yield a much smaller
 value for $m_{J/\Psi}-m_{\eta_c}$ when compared with the true experimental value.
 In our unquenched study, however, we see that this difference is
 somewhat milder compared with the previous quenched situation.
 Both our lattice result and the previous quenched result for
 this quantity  are to be compared with the value
 $\Gamma_{PDG}=1.58(38)$ KeV quoted by the PDG.
 Note that the PDG value is an average of CLEO result
 and the Crystal Ball result, the former being
 $1.92(30)$ KeV which is closer to our lattice result
 while the latter from Crystal Ball being $1.18(33)$ KeV, smaller
 than lattice results.

 \subsection{$\chi_{c0}\to J/\Psi\gamma$ transition}

 In the continuum, this transition matrix element has
 the following decomposition:~\cite{dudek:06}
 \begin{widetext}
 \ba \label{eq:chic0_decomposition} \langle
 S(\bp_S)|j^\mu(0)|V(\bp_V,r)\rangle&=&\Omega^{-1}(Q^2)\Bigg(
 E_1(Q^2)[\Omega(Q^2)\epsilon^\mu(\bp_V,r)-\epsilon(\bp_V,r)\cdot p_S(p_V^\mu p_V\cdot p_S-m_V^2p_S^\mu)]\nonumber\\
 & & +\frac{C_1(Q^2)}{\sqrt{q^2}}m_V\epsilon(\bp_V,r)\cdot p_S[p_V\cdot p_S(p_V+p_S)^\mu-m_S^2p_V^\mu-m_V^2p_S^\mu]\Bigg)
 \ea
 \end{widetext}
 with $\Omega(Q^2)=\left(p_V\cdot p_S\right)^2-m_V^2m_S^2$.
 Therefore, the hadronic matrix element is characterized by
 two form factors $E_1(Q^2)$ and $C_1(Q^2)$.
 At the physical photon point with $Q^2=0$, only the former
 contributes.

\begin{figure*}[h]
 \centering
 \includegraphics[scale=0.8]{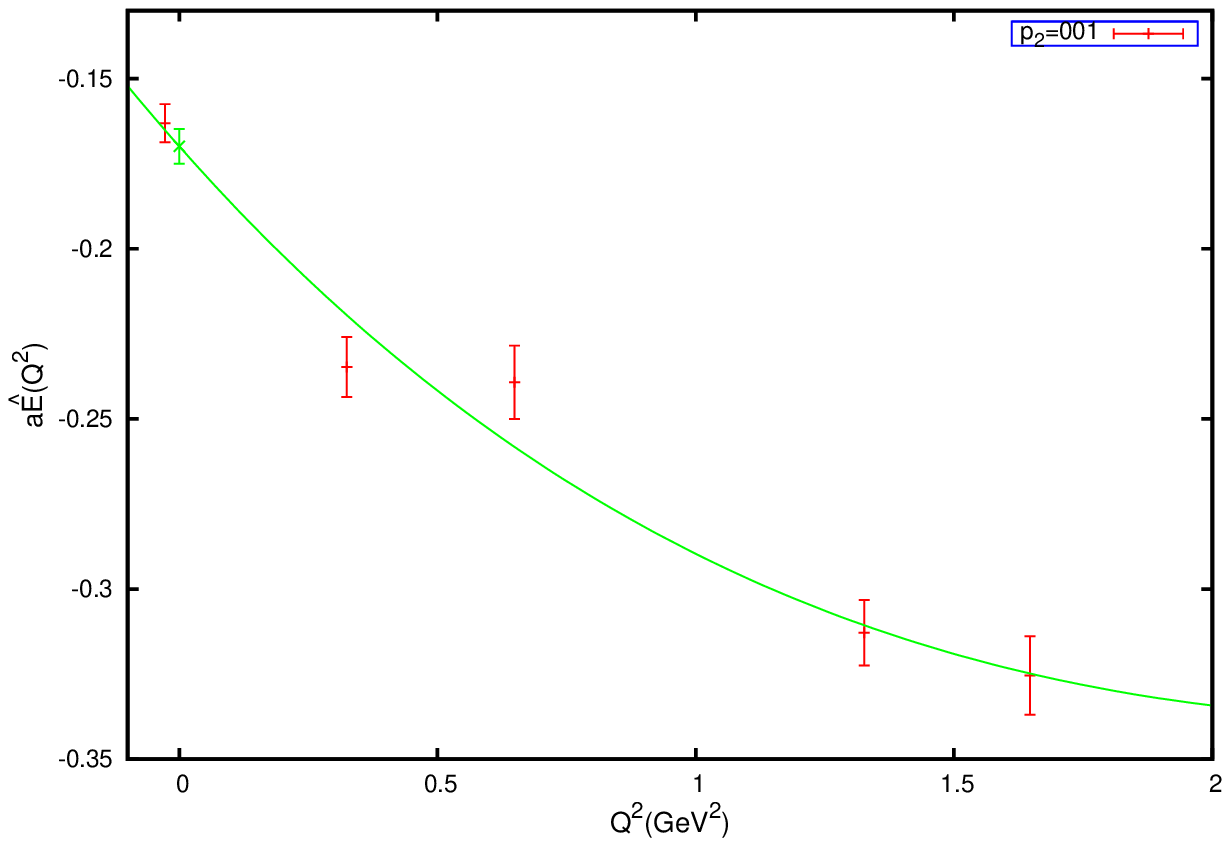}
 \caption{$\chi_{c0}\to J/\Psi \gamma$ transition form factor
 $\hat{E}_1(Q^2)$ obtained from our simulation. The curve is
 a fit according to Eq.~(\ref{eq:chic0-to-jpsi-fit}).
 The fitted value of $\hat{E}_1(Q^2=0)$ is also indicated
 at $Q^2=0$ with its error.}
 \label{fig:chic0_to_jpsi}
 \end{figure*}

 The form factor $E_1(Q^2)$ can be obtained by following a similar process
 as the other form factors.
 We can always choose some combinations of $p_V,p_S$ such that
 \[
 \langle S(\bp_S)|j^\mu(0)|V(\bp_V,r)\rangle\propto E_1(Q^2).
 \]
 The final lattice results for $\hat{E}_1(Q^2)$ are
 shown in Fig.~\ref{fig:chic0_to_jpsi}.
 We then use the following form
 \be
 \label{eq:chic0-to-jpsi-fit}
 \hat{E}_1(Q^2)=\hat{E}_1(0)\left(1+\frac{Q^2}{\rho^2}\right)\exp\left[-\frac{Q^2}{16\beta^2}\right],
 \ee
 to fit the data~\cite{dudek:06}.
 The fitted parameters we obtain are:
 \ba
 a\hat{E}_1(0) &=& -0.1699(51)\;,
 \\
 \rho &=& 871(85)~\mbox{MeV},\, \beta = 451(62)~ \mbox{MeV}
 \;.\nonumber
 \ea
 The fitted value of $\hat{E}_1(0)$ at $Q^2=0$ is also
 indicated in Fig.~\ref{fig:chic0_to_jpsi} together with
 its error. These results are to be compared with similar results
 from previous quenched lattice study~\cite{dudek:06}:
 $a_t\hat{E}_1(0)= -0.137(12)$, $\beta=542(35)\mbox{MeV}$
 and $\rho=1.08(13)~\mbox{GeV}$.

 At the physical photon point $Q^2=0$, the decay width
 for this radiative transition is given by:
 \be
 \label{eq:chic0_decay_width}
 \Gamma\left(\chi_{c0}\to  J/\Psi\gamma\right)=\alpha\frac{16}{9}\frac{|\bq|}{m^2_{\chi_{c0}}}|\hat{E}_1(0)|^2
 \ee
 where $\hat{E}_1$ is related to $E_1$ by:
 \[
 E_1(Q^2)=2\times\frac23e\times\hat{E}_1(Q^2)
 \]
 Substituting our lattice result for $\hat{E}_1(0)$,
 we then can get the decay width in physical unit:
 \be
 \Gamma_{m_{\rm phy}}=85(7)~\mbox{KeV},\,\Gamma_{m_{\rm lat}}=65(4)~\mbox{KeV}
 \ee
 which is to be compared with the quenched lattice
 result of $\Gamma_{m_{\rm phy}}=232(41)~\mbox{KeV}$ and
 $\Gamma_{m_{\rm lat}}=288(60)~\mbox{KeV}$. It is seen
 that our unquenched result for this decay width is
 substantially smaller than their quenched values.
 The value for this quantity quoted by the PDG is
 given by:
 $\Gamma_{PDG}=119(11)~\mbox{KeV}$
 which lies in between the quenched and unquenched
 results.

\subsection{$h_c\to \eta_c\gamma$ transition}

 The form factor decomposition for this process
 is identical to Eq.~(\ref{eq:chic0_decomposition}).
 However, the signal for the state $h_c$ is much noisier.
 It turns out that we could only get reasonable signal with $\bp_{h_c}=000$.
 In order to get various values of $Q^2$,
 we vary the values of $\bq$ and $\bp_{\eta_c}$ simultaneously
 such that $h_c$ is always at rest.

\begin{figure*}[h]
 \centering
 \includegraphics[scale=0.8]{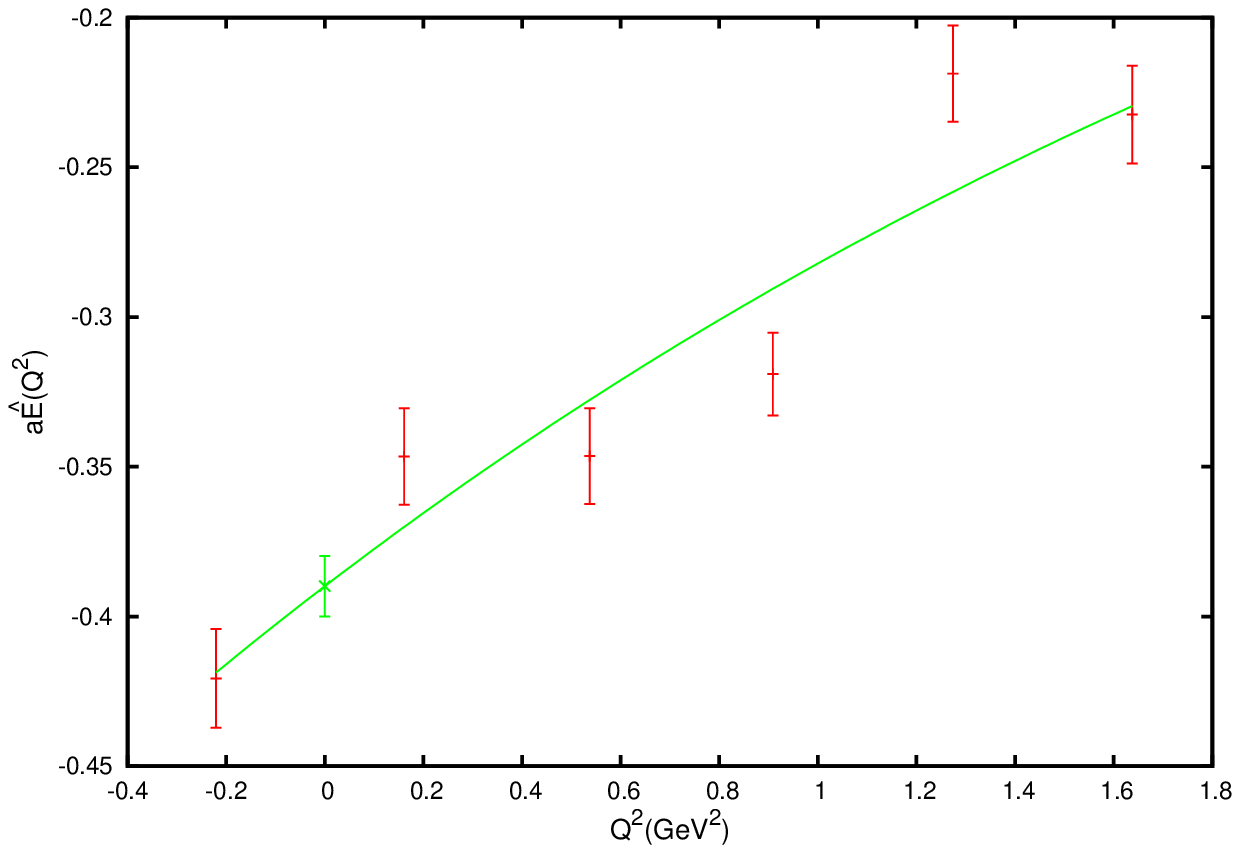}
 \caption{The $h_c\to \eta_c \gamma$ transition form factor
 $\hat{E}_1(Q^2)$ obtained from our calculation (the data points).
 The curve is a fit using Eq.~(\ref{eq:hc-fit}). Also
 shown at $Q^2=0$ is the fitted value
 of $\hat{E}_1(0)$ together with its error.}
 \label{fig:hc_to_etac}
 \end{figure*}

 The form factor we obtain is illustrated in Fig.~\ref{fig:hc_to_etac}.
 We fit the data with a functional form:~\cite{dudek:06}
 \be
 \label{eq:hc-fit}
 \hat{E}_1(Q^2)=\hat{E}_1(0)\exp\left[-\frac{Q^2}{16\beta^2}\right]
 \;,
 \ee
 and the fitted parameters come out to be:
 \be
 a\hat{E}_1(0)= -0.39(1)   ,\,    \beta = 440(23)~\mbox{MeV}
 \ee
 The fitted value of $\hat{E}_1(0)$ is also shown in
 Fig.~\ref{fig:hc_to_etac} at $Q^2=0$ together with
 the corresponding error.
 These numbers are to be compared with the corresponding
 quenched results: $a_t\hat{E}_1(0)= -0.306(14)$
 and $\beta=689(133)~\mbox{MeV}$~\cite{dudek:06}.

 The physical decay width for the transition is given by:
 \be
 \Gamma\left(h_c \to \eta_c\gamma\right)=\alpha\frac{16}{27}\frac{|\bq|}{m^2_{h_c}}|\hat{E}_1(0)|^2
 \ee
 With our lattice result for $\hat{E}_1(0)$ substituted in, we find:
 \be
 \Gamma_{m_{\rm phy}}=234(12)~\mbox{KeV},\,\Gamma_{m_{\rm lat}}=210(13)~\mbox{KeV}
 \;.
 \ee
 The corresponding quenched lattice values are:
 $\Gamma_{m_{\rm phy}}=601(55)~\mbox{KeV}$ and
 $\Gamma_{m_{\rm lat}}=663(132)~\mbox{KeV}$, both
 of which are about a factor of 3 larger than our
 unquenched result, though the errors are somewhat large.
 The lattice results for this decay  can now be
 compared with the recent measurement at BES-III.~\cite{BESIII}
 The total width for $h_c$ and the corresponding
 branching ratio for the radiative transition are found to be
 \ba
 \Gamma^{\rm tot}_{\rm exp} &=&730\pm450\pm280~\mbox{KeV}\;,
 \nonumber \\
 B(h_c\to \eta_c\gamma) &=& (54.3\pm6.7\pm5.2)\%\;,
 \nonumber
 \ea
 where the first error is statistical and the second is
 systematic. If we multiply the central values for the above
 two quantities and add the errors in quadrature we find
 the decay width $\Gamma(h_c\to\eta_c\gamma)= 396(294)$keV,
 which could be compared with our lattice result.
 The agreement within a large error is seen although improvements from
 both experiment and lattice calculations are required
 to cut down the large uncertainties for this quantity.

\section{Summary and conclusions}
\label{sec:conclude}

 \begin{table*}[ht]
 \centering \caption{Summary of the results obtained in this work. Also listed
 are the corresponding results from quenched lattice QCD~\cite{dudek:06}. Experimental
 values or values from PDG are also listed whenever available.}
 \label{tab:summary}
 \begin{tabular*}{8cm}{l|p{3cm}|p{3cm}}
 \hline
   \multicolumn{3}{c}{Fitted parameter $\beta$[MeV] for form factors}\\
    \hline
    & $\eta_c$ & $\chi_{c0}$ \\
    \hline
     This work   & 567(2) &  510(16)\\
 Ref.~\cite{dudek:06} & 480(3) & 393(12)\\
 \hline
 \hline
 \end{tabular*}
 \begin{tabular*}{9cm}{l|l|l|l}
 \hline
    & \multicolumn{3}{c}{$\Gamma_{m_{\rm phy}}/\Gamma_{m_{\rm lat}}$[keV] for transitions}\\
 \cline{2-4}
 & $J/\Psi\to\eta_c\gamma$ & $\chi_{c0}\to J/\Psi\gamma$ & $h_c\to\eta_c\gamma$ \\
 \hline
  PDG   &  1.58(38)     & 119(11)    & 396(294)\\
 This work                        &  2.84(6)/1.99(6) & 85(7)/65(4)     & 234(12)/210(13)  \\
 Ref.~\cite{dudek:06}  &  2.57(11)/1.61(7)& 232(41)/288(60) & 601(55)/663(132) \\
 \hline
 \end{tabular*}
 \end{table*}
 In this exploratory study, we calculate the form factors for some
 of the ground state charmonia and their radiative transitions using
 unquenched $N_f=2$ twisted mass fermions. The $s$ and $c$ quarks are
 quenched which are incorporated via a twisted doublet for each non-degenerate
 quark flavor in the valence sector.
 Our study focuses on the form factors for $\eta_c$, $\chi_{c0}$ and the
 $J/\Psi\to \eta_c\gamma$, $\chi_{c0}\to J/\Psi\gamma$, $h_c\to \eta_c\gamma$
 radiative transitions. The mass spectrum and dispersion relations
 for these charmonium states are first examined. Good agreement of
 the computed spectrum with the experiment is found. It is also
 verified that, by using lattice dispersion relations instead of
 the naive continuum ones, the lattice artifacts for these
 charmonium states are well under control.
 By computing various appropriate ratios
 of the three-point functions to the two-point functions, hadronic matrix elements
 for these transitions are obtained at various of $Q^2$.
 Using the parameterized form in terms of relevant form factors,
 we obtain the lattice results for the relevant form factors
 and the radiative decay widths
 for these channels. Our results are summarized in
 Table~\ref{tab:summary} which are to be compared with those obtained
 in previous quenched lattice studies and experimental values.

 Although some quantities
 from our unquenched study turn out to be comparable with
 the quenched results, quite a number of the results still differ substantially,
 as is seen from Table~\ref{tab:summary}.
 For example, for the form factors of $\eta_c$ and $\chi_{c0}$
 a  harder behavior (a larger value of $\beta$ hence a smaller charge radius) than the quenched
 result is found. As for the decay width for $J/\Psi\to
 \eta_c\gamma$, a value larger than the quenched result is
 obtained. Due to the improvement of the mass splitting
 for the charmonium in unquenched study, the discrepancy
 between the results using the physical mass and the lattice
 computed mass is somewhat narrowed, although the value is still
 larger than the value quoted by PDG.
 For the decay width of $\chi_{c0}\to J/\Psi\gamma$, a value
 smaller than the PDG value (and also the quenched result) is obtained.
 As for the $h_c\to \eta_c\gamma$ transition,
 the signal is noisy and our unquenched result is much smaller
 than the quenched value with a large statistical uncertainty.
 It is  still compatible with the recently measured value at BES-III
 which also has a large error. To get better signals for
 this channel, variational methods or smearing techniques might be
 necessary which will be investigated in the future.

 In this preliminary study, we simulate at only one lattice spacing and sea quark mass,
 and no chiral nor continuum extrapolation is made.
 The physics involved in this study mainly concerns the heavy flavor part of
 the theory which should not be sensitive to the pion mass.
 As for the lattice artifacts, we argued that, thanks to the automatic $\calO(a)$ improvement, the
 lattice artifacts is under control. Indeed, by using the lattice dispersion
 relations, we verified that all charmonium states that
 we studied exhibit controlled lattice errors in their dispersion relations
 of about a few percent, which
 is roughly at the order of $(\mu_ca)^2$ for our simulation.
 With the experience gained in this study, it would be better and
 also possible to study charmonium radiative transitions in a more systematic
 manner (more lattice spacings, more pion mass values etc.) using unquenched
 lattice QCD. It is also tempting to perform similar studies with the $N_f=2+1+1$
 dynamical twisted mass fermion. Given the promising
 experimental status of BES-III at BEPC-II, the unquenched lattice studies
 on charmonium transitions will certainly be an interesting project to pursue
 in the future.

\section*{Acknowledgments}
 The numerical computations  for this project was performed on the
 Magic Cube at Shanghai Supercomputer Center and on Tianhe-1A at
 National Supercomputing Center in Tianjin. We thank the ETMC for
 allowing us to use their gauge field configurations and part of
 their packages. We thank Dr. X.~Feng, K.~Jansen and M.~Wagner for
 valuable discussions. This work is supported in part by the
 National Science Foundation of Chian (NSFC) under project
 No.10835002, No.11021092, No.10675101, No.11075167 and No.10975076.


\end{document}